\documentclass{proc-l}

\usepackage{amssymb}

\usepackage{graphicx}

\newtheorem{theorem}{Theorem}[section]
\newtheorem{lemma}[theorem]{Lemma}

\theoremstyle{definition}
\newtheorem{definition}[theorem]{Definition}

\theoremstyle{remark}

\numberwithin{equation}{section}

\begin{document}

\title{The Finiteness of vortices in steady incompressible viscous fluid flow}


\author{Jiten C. Kalita}
\address{Department of Mathematics, Indian Institute of Technology Guwahati, Guwahati, India 781039}
\email{jiten@iitg.ernet.in}

\author{Sougata Biswas}
\address{Department of Mathematics, Indian Institute of Technology Guwahati, Guwahati, India 781039}
\email{b.sougata@iitg.ernet.in}
\author{Swapnendu Panda}
\address{Department of Mathematics, Indian Institute of Technology Guwahati, Guwahati, India 781039}
\email{p.swapnendu@iitg.ernet.in}


\date{}

\dedicatory{}

\commby{}

\begin{abstract}
In this work, we provide two novel approaches to show that incompressible fluid flow in a finite domain contains at most a finite number vortices. We use a recently developed geometric theory of incompressible viscous flows along with an existing mathematical analysis concept to establish the finiteness.  We also offer a second proof of finiteness by roping in the Kolmogorov's length scale criterion in conjunction with the notion of diametric disks.
\end{abstract}

\maketitle
\section{Introduction}
The study of a general system of vortices generated due to flow separation in incompressible viscous flows in a finite domain has eluded proper investigation for quite a long period of time. Most of the available literatures are concerned about flow in some nice and standard geometry only, such as Moffatt vortices (sequence of vortices present in the solid corners of internal viscous incompressible flows) \cite{bk,kirk,kras1,malyuga,moff1,shankar1,shtern}, vortices in a semi-circular domain \cite{glow} etc. The finiteness of such system in the general scenario is still wide open. In the present work we propose a couple of rigorous proofs in the direction of finiteness of such an arbitrary system of vortices.

Studies on the finiteness of critical  points in a flow domain found through dynamical system approach can be found scantily in the existing literature \cite{wang,wang2}. However, the connection between these critical points and the finiteness of vortices  in a bounded domain has yet to catch the attention of the scientific community. To the best of our knowledge, no rigorous mathematical proof has been provided to establish this connection.
In order to achieve this, we utilize some ideas from the geometric theories of incompressible viscous flows along with the Kolmogorov's length scale criterion \cite{deshpande,wal}.

Kolmogorov proposed his famous 1st similarity hypothesis in the year 1941 \cite{kolmo}. He propounded that the only factors influencing the behavior of the small scale motions are the overall kinetic energy production rate (which equals the dissipation rate) and the viscosity \cite{andersson, fris, jim}. While the dissipation rate is independent of viscosity, the scales at which this energy is dissipated depends both on the dissipation rate and viscosity. If the dissipation rate per unit mass ($\varepsilon$) has  dimensions ($m^2/sec^3$) and viscosity, $\nu$ has dimension ($m^2/sec$) then the length scale formed from these quantities is given by
$$\eta=\bigg(\frac{\nu^3}{\varepsilon}\bigg)^{1/4}$$
This length scale is the so called Kolmogorov length scale, beyond which size eddies cannot be formed either in laminar or turbulent flow \cite{deshpande,wal}. Otherwise, the local value of power density ($\varepsilon$) would be so high that the kinetic energy would be fully dissipated as heat.

We provide two proofs which establish the finiteness of vortices in a bounded domain, one using the Bolzano-Weierstrass theorem \cite{rudin} and the other one using the diametric disk concept \cite{jiten}.
\section{Preliminaries}
\subsection{Notations}
\begin{itemize}
\item $\mathbb{R}$ is the set of real numbers.
\item $M$ is a planar region.
\item On the boundary $\partial M$ of the region $M$, $\hat{\tau}$ denotes the tangential vector and $\hat{n}$ the normal vector.
\item $T_pM = \{w\;|\: w \;\mbox{is tangent to} \; M\;\mbox{at}\; p\}$.
\item $TM = \{(p, T_pM)\;|\;p \in M\}$ is the tangent bundle of $M$.
\item Assume $r \geq 1$ is an integer. Let $\mathcal{C}^r(TM)$ be the space of all $r$-th differentiable vector fields $v$ on $M$.
\item $\mathcal{C}^r_{\hat{n}}(TM)= \{v \in \mathcal{C}^r(TM)\;|\;v\cdot \hat{n} = 0$ on $\partial M\}.$
\item $\mathcal{D}^r(TM)= \{v \in \mathcal{C}^r_{\hat{n}}(TM)\;|\;\nabla\cdot v = 0\}.$
\item $\mathcal{B}_0^r(TM)= \{v \in \mathcal{D}^r(TM)\;|\;v = 0$ on $\partial M\}.$
\end{itemize}
\subsection{Geometric theory of viscous incompressible flows}
\begin{definition}\cite{wang,wang2}
\emph{A point $p \in M$ is called a singular point of $v \in \mathcal{C}^r_{\hat{n}}(TM)$ if $v(p)=0$.}
\end{definition}
\begin{definition}\cite{wang,wang2}
\emph{A singular point $p$ of $v \in \mathcal{C}^r_{\hat{n}}(TM)$ is called non-degenerate if the Jacobian matrix of $v$ at $p$, $J_v(p)$ is invertible.}
\end{definition}
\begin{definition}\cite{wang,wang2}
A vector field $v \in \mathcal{C}^r_{\hat{n}}(TM)$ is called regular if all singular points of $v$ are non-degenerate.
\end{definition}
\begin{definition}\cite{wang,wang2}
\emph{
Let $v \in \mathcal{B}^r_0(TM) (r \geq 2).$
\begin{enumerate}
\item[(1)] A point $p \in \partial M$ is called a $\partial$-regular point of $v$ if $\displaystyle \frac{\partial v_{\hat{\tau}}(p)}{\partial n} \neq 0$; otherwise, $p \in \partial M$ is called a $\partial$-singular point of $v$.
\item[(2)] A $\partial$-singular point $p \in \partial M$ of $v$ is called non-degenerate if
$$\displaystyle \mathrm{det}\begin{bmatrix}
\frac{\partial^2v_{\hat{\tau}}(p)}{\partial \hat{\tau}^2}
 &\frac{\partial^2v_{\hat{\tau}}(p)}{\partial {\hat{\tau}} \partial \hat{n}} \\
\frac{\partial^2v_{\hat{n}}(p)}{\partial \hat{\tau} \partial \hat{n}} & \frac{\partial^2v_{\hat{n}}(p)}{\partial \hat{n}^2}\\
\end{bmatrix} \neq 0
$$
A non-degenerate $\partial$-singular point of $v$ is also called a $\partial$-saddle point of $v$.
\end{enumerate}
}
\end{definition}
\begin{lemma}\label{lm_bd}
{\bf \cite{wang,wang2}} Each non-degenerate $\partial$-singular point of $v \in \mathcal{B}_0^r(TM)$ is isolated. 
\end{lemma}
\section{Finiteness of vortices}
In this section we will establish the finiteness of an arbitrary system of vortices (in incompressible viscous flows) through two different approaches. The first proof has its origin in mathematical analysis \cite{rudin} and recently developed geometric theory of incompressible viscous flows \cite{wang,wang2}. The second proof leans more on the geometric aspects of the vortices which incorporates of the concept of diametric disk.
\begin{theorem}
Suppose $\Omega$ is a closed subset of $\mathbb{R}^2$, representing an enclosed domain bounded by solid walls (or combination of solid walls and free surfaces). Then for a steady incompressible viscous flow field defined by the velocity vector $v$, for every point $p$ on the boundary, any neighbourhood of $p$ contains at most a finite number of vortices.
\end{theorem}
\begin{itemize}
\item[{\it Proof 1:}]
It is well known that the process of flow separation on the solid boundaries of a  domain leads to the creation of vortices in the flow field. The points on the boundary where this flow separation takes place, are termed as separation points. This class of critical points follow the definition of  non-degenerate $\partial$-singular points \cite{wang,wang2}. \\

If possible, suppose number of such critical points is infinite. Let us denote the set of separation points by $\mathfrak{C}$. But since the flow domain ($\subset \mathbb{R}^2$) is always bounded, and all these critical points are points from the solid boundary of the flow domain, hence $\mathfrak{C}$ is an infinite bounded subset of $\mathbb{R}^2$. By Bolzano-Weierstrass theorem \cite{rudin}, $\mathfrak{C}$ must have a limit point ($c_0$ say) in $\mathbb{R}^2$. Therefore, there exists a sequence $(c_k) \subset \mathfrak{C}$ such that $c_k \to c_0$ as $k \to \infty$. Moreover, since flow separation takes place in presence of a solid boundary,  this point $c_0$ has to lie in the flow domain on the solid boundary. Also, the flow field is continuously differentiable, hence a convergent sequence of $\partial$-singular points must converge to a $\partial$-singular point. This contradicts the fact that $\partial$-critical points are isolated (Lemma \ref{lm_bd}). Therefore, number of separation points must be finite.\\

Since, separation points lead to the creation of vortices, the finiteness of separation points leads to the finiteness of vortices (see the note at the end of Proof 2) in the flow domain.\\
\hfill $\blacksquare$
\item[{\it Proof 2:}]
Let $V$ be the set representing a vortex. Define,
$$\mbox{diam}(V) = \min_{p \in \partial V} \{ 2 \Vert c_v-p \Vert_2 : c_v \mbox{ is the center of the vortex } V \}.$$
Consider the disk of radius $\frac{\mbox{diam}(V)}{2}$ centered at $c_v$. This is the largest disk inscribed in the vortex $V$ with center at $c_v$. We call them the diametric disk of $V$ \cite{jiten}. Refer to Figure \ref{eq1} for a schematic of diametric disk in the lid-driven cavity flow \cite{bk}.
\begin{figure}[hH]
\begin{center}
\includegraphics[height=7.5cm]{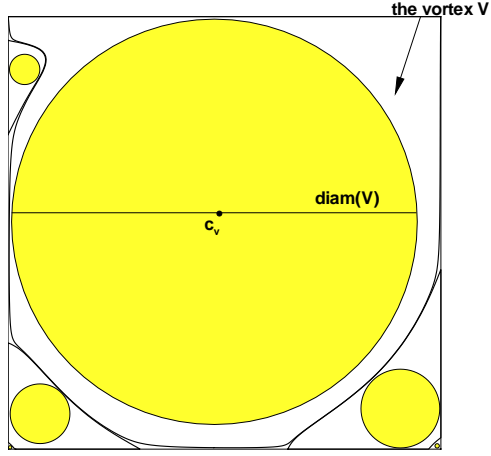}
\caption{Schematic of diametric disk in the lid-driven cavity flow.}\label{eq1}
\end{center}
\end{figure}

The total area of the flow domain covered by $V$ must be $\geq \frac{\pi}{4}$ $\mbox{diam}^2(V)$. Moreover since the vortices are non-overlapping, hence the diametric disks are mutually disjoint. Therefore, the total area covered by the vortices must be $\geq$ the sum of the areas of the diametric disks, i.e.
$$Area (\mbox{flow domain}) > \sum_{k\in \Lambda} Area(V_k) \geq \frac{\pi}{4} \sum_{k\in \Lambda} \mbox{diam}^2(V_k), $$
where $\Lambda$ is an index set. But since none of the diameters can drop below the Kolmogorov length scale \cite{wal}, therefore
$$ \sum_{k\in \Lambda} \mbox{diam}^2(V_k) < \infty \mbox{ only if } \vert \Lambda \vert < \infty.$$
As the total area of the flow domain is itself finite, hence the result follows.
\hfill $\blacksquare$\\
\par
\noindent
Therefore, the finiteness of the system of vortices is indeed independent of the geometry of the flow domain. The second proof suggests that even if the flow domain is not simply connected (contains holes), still the finiteness of number of vortices is guaranteed.
\end{itemize}
{\bf NOTE:}\\
The separation points on the solid boundary paves the way for the formation of a new vortex. But one separation point ($\partial$-singular point) does not necessarily generate only a single vortex in general over a period of time. In fact, in the event of  breakdown of vortices during vortex-shedding for flow past bluff bodies, at a particular instant of time the set of vortices generated from the same separation point is still  finite. The instantaneous vortex will be attached to the solid boundary, whereas the earlier ones will be pushed into the downstream of the flow domain. The second proof of finiteness of vortices ensures the finiteness of vortices corresponding to the same separation point. So the set of all vortices in the flow domain being a finite union of finite sets, is finite.
\section{Conclusion}
The geometric theories of incompressible viscous flows as a discipline of Topological Fluid Dynamics mainly revolve around the existence of critical points in a flow field. To the best of our knowledge, no connection of these critical points to the number of vortices in a flow field has been established so far. 
We provide two rigorous mathematical proofs to show that the number of vortices in an incompressible fluid flow in a finite domain is finite. For the first proof, a recently developed geometric theory of incompressible viscous flows along with an existing mathematical analysis concept has been used. The second proof adopts the Kolmogorov's length scale criterion in conjunction with the notion of diametric disks.

\bibliographystyle{amsplain}

\end{document}